\documentclass[twocolumn,showpacs,preprintnumbers,amsmath,amssymb,aps,prb,floatfix]{revtex4}
\usepackage{graphicx}
\usepackage{times}
\usepackage{subfigure}
\usepackage{epsfig}
\newcommand{\V}{\boldsymbol}

\begin{document}

\title{Polaron relaxation in self-assembled quantum dots: Breakdown of the semi-classical model\\}

\author{T. Grange}
\author{R. Ferreira}
\author{G. Bastard}

\affiliation{Laboratoire Pierre Aigrain, Ecole Normale
Sup\'erieure, 24 rue Lhomond, 75005, Paris, France}

\begin{abstract}

We calculate the lifetime of conduction band excited states in self-assembled quantum dots by taking into account LO-phonon-electron interaction and various anharmonic phonon couplings.  We show that polaron relaxation cannot be accurately described by a semi-classical model. The contributions of different anharmonic decay channels are shown to depend strongly on the polaron energy. We calculate the energy dependence of polaron lifetime and compare our results to available experimental measurements of polaron decay time in InAs/GaAs quantum dots.
\end{abstract}

\pacs{73.21.La,71.38.-k,76.60.Es}

\maketitle

Carrier relaxation in quantum dot (QD) is of fundamental importance for potential applications ($e.g.$ laser, photo-detectors...).
In self-assembled quantum dots, the energy separation between the ground and first excited states is of the order of tens of meV.
Early theoretical studies of carrier relaxation in these nanostructures led to the so-called phonon bottleneck, in which the phonon-assisted relaxation is strongly inhibited  because of the discreteness of electronic states\cite{bockelmann90,benisty91,inoshita92}. Further studies predicted the existence of a strong coupling regime between electron and longitudinal optical (LO) phonons in self-assembled QDs \cite{inoshita97}.
This strong coupling regime leads to the formation of polarons, which are entangled electron-LO-phonon quasi-particles.  Polaron states were first observed in intra-band magneto-transmission experiments done on n-doped QDs \cite{hameau99,hameau02,carpenter06}. More recently, evidence of a strong coupling regime has also been obtained for holes\cite{preisler05} and electron-hole pairs\cite{preisler06} confined in QDs.
A striking consequence of the polaron formation is that electrons can no longer relax by LO-phonon emission.  Nevertheless, efficient polaron relaxation (lifetime $\tau_{pol}$ of few tens of picosecond) has been measured in QDs doped with one electron\cite{sauvage02,zibik04} and more recently with two electrons \cite{grange07}.

So far, there are two models in the literature to describe the polaron relaxation in QDs. Li \textit{et al} \cite{li99} were the first to consider the carrier relaxation as triggered by the intrinsic (bulk related) instability of LO-phonons.  They discussed a semi-classical model in which a phonon damping was phenomenologically added when solving the time-dependent Schrodinger equation for the coupled electron-LO phonon system.  They were able to derive a formula for the polaron lifetime, which depends on two parameters: the strength of the electron-LO-phonon coupling and the damping rate.  The later were taken as fit parameters in Refs~\onlinecite{sauvage02} and \onlinecite{zibik04} to successfully interpret the increase of $\tau_{pol}$ with the polaron energy $E_{pol}$ in the 40-52 meV interval.  However, this model fails in explaining the decrease of $\tau_{pol}$ for polarons with higher energy\cite{zibik04} (up to $\sim 60$ meV). On the other hand, Verzelen \textit{et al}\cite{verzelen00} and Jacak \textit{et al}\cite{jacak02} have applied the Fermi golden rule to polaron states, assuming the anharmonic mechanism proposed by Vallee and Bogani for bulk LO-phonons (LO $\rightarrow$ LO + TA) \cite{vallee91}. This assumption leads to the existence of a narrow energy window for relaxation : only polarons with an energy $E_{pol}$ in the $\sim$ 35-44 meV interval can disintegrate.  This model also predicts an increase of $\tau_{pol}$ with $E_{pol} > \hbar\omega_{LO}$ (where $\omega_{LO}$ is the LO-phonon frequency), but of course fails when applied to higher energy polarons. 

In this paper we present a detailed study of polaron relaxation in QDs.  We calculate the lifetime of excited states in self-assembled QDs by fully taking into account their polaronic nature as well as the phonon anharmonicity.  We show that the variation of $\tau_{pol}$ with $E_{pol}$ is not governed by the sole weight of its LO-phonon component (like in the work of Li \textit{et al}), neither can be restricted to only one bulk disintegration channel (like in the works of Verzelen \textit{et al} and Jacak \textit{et al}), but that different anharmonic channels have to be taken into account depending on $E_{pol}$. As we show below, the polaron lifetime is equal to the product of its LO-phonon weight by the decay rate of a bulk LO-phonon \textit{which would have the polaron energy}.  We obtain a good overall agreement with the experimental data for any polaron energy.  In particular, the model allows to explain the measured unexpected decrease of $\tau_{pol}$ with increasing $E_{pol}$ at high energies.

We consider a quantum dot hosting one electron in its conduction band (as obtained \textit{e.g.} by doping).
The system is modelled by the following Hamiltonian:
\begin{equation}
H = H_e + H_{vib} + H_{e-ph}
\end{equation}
where $H_e $ is the purely electronic Hamiltonian, $H_{vib}$ the vibrational one and $H_{e-ph}$ represents
the different electron-phonon interactions.
The vibrational Hamiltonian (assumed to be bulk-like), can be expressed as $H_{vib} = H_{ph}+ V_{a}$
where $H_{ph}$ is the harmonic hamiltonian of  phonons while $V_a$ is the anharmonic part.
In actual self-assembled QDs, the energy spacing between ground and first excited level is much larger than the energy of acoustic phonons which have the QD size wavelength. Therefore, we can neglect direct couplings related to acoustic phonons and consider only the Fr\"ohlich Hamiltonian:
\begin{equation}
H_f=\sum_{\V{q}}\frac{i C_ f}{q} \frac{e^{i\V{q}.\V{r}}}{\sqrt{V_{cr}}}
(a_{\V{q}}-a_{\V{-q}}^{+})
\end{equation}
where
$C_{f} = e\sqrt{\frac{\hbar}{2\varepsilon_0\varepsilon_{\infty}\omega_{LO}}(\omega_{LO}^2-\omega_{TO}^2)}$, $\omega_{LO}$ and $\omega_{TO}$ are respectively the zone-center LO and TO modes frequencies, $\varepsilon_{\infty}$ is the high frequency permittivity and $V_{cr}$ is the crystal volume.

Let us define $H_0 = H_e + H_{ph} +H_f $. The polaron states are eigensolutions of $H_0$. We recall briefly in the following their main aspects (for more details on polaron eigenstates see e.g. Ref~\onlinecite{hameau02}).
The polaron wavefunctions are entanglements
of decoupled states $|a,n\rangle$ where $a=s,p_{x}$ or
$p_{y}$ is the electron wavefunction (solution of $H_e$) and $n$ is the number of LO phonon modes (solutions of $H_{ph}$)
entering in the formation of the polaron.
Since the LO-phonons that are involved in the Fr\"ohlich couplings have small wavevectors \cite{inoshita97},  we can assume that LO phonon are dispersionless.
Within this approximation, the
Fr\"ohlich interaction between the discrete zero phonon level
$|p,0\rangle$ and the one phonon flat continuum $|s,1_{\V{q}} \rangle$ can be
treated as an interaction between two discrete levels, namely
$|p,0\rangle$ and $|s,1_{sp}\rangle$, where
the phonon mode $\vert 1_{sp} \rangle$ is defined by:
\begin{equation}
\vert 1_{sp} \rangle = \sum_{q} \dfrac{\langle s,1_{\V{q}} \vert H_f \vert p,0 \rangle}{V_{sp}} \vert 1_{\V{q}} \rangle
\end{equation}
where $|1_{\V{q}} \rangle$ are bulk LO-phonon states, and $V_{sp}=\sqrt{\sum_{q} \vert\langle s,1_{\V{q}} \vert H_f \vert p,0 \rangle\vert^2}$ is the coupling strength between $|p,0\rangle$ and $\vert s,1_{sp} \rangle$.
\begin{figure}
\centering
\mbox{
\subfigure[]{\epsfig{figure=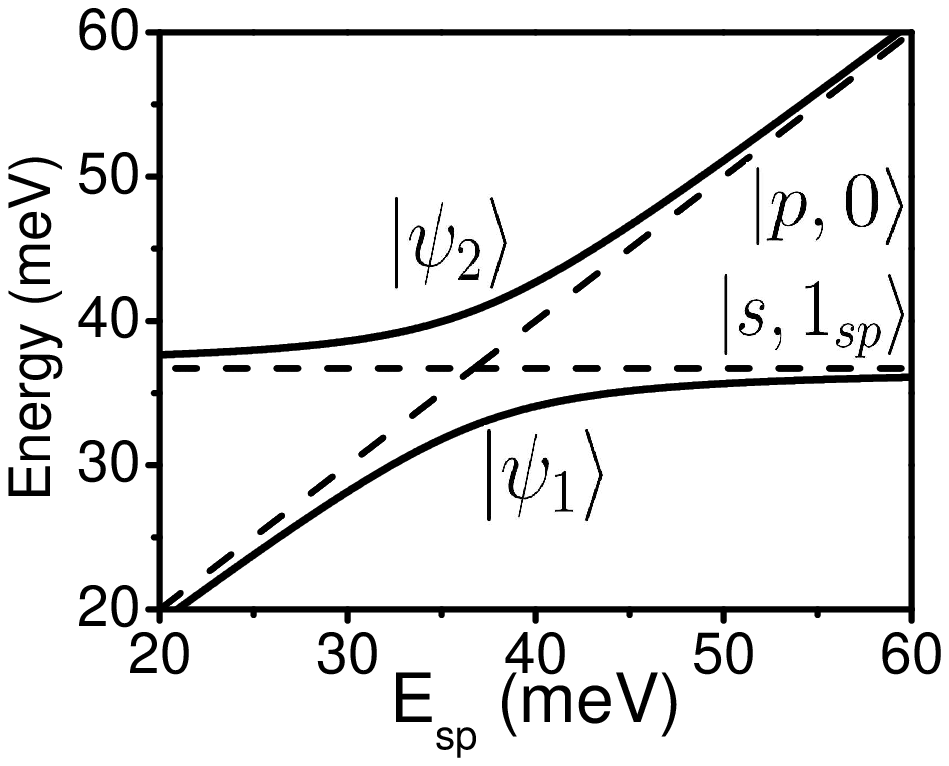,width=0.23\textwidth}} \quad
\subfigure[]{\epsfig{figure=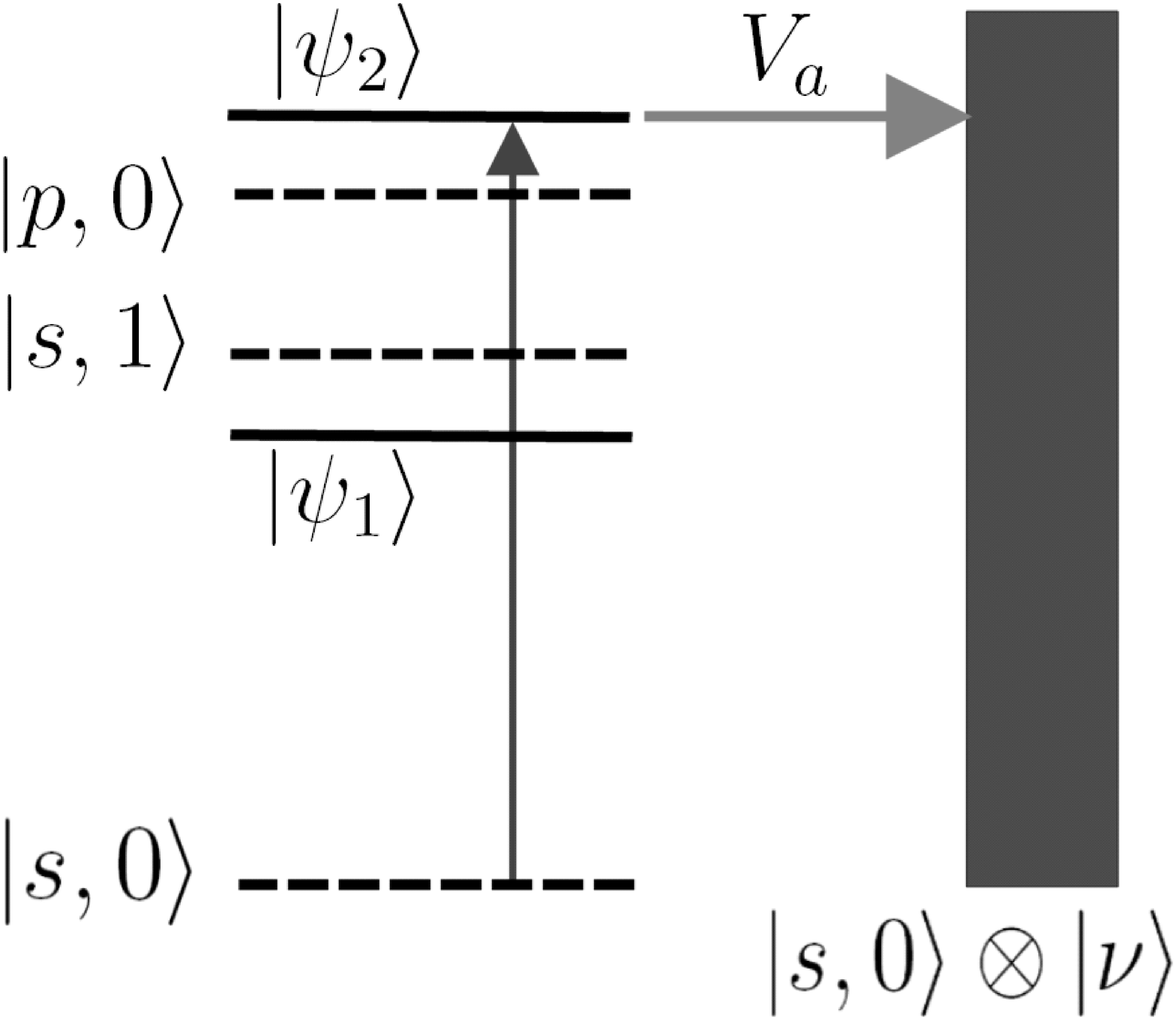,width=0.23\textwidth}}}
\caption{\footnotesize (a) Full lines: Polaron energies as a function of the electronic energy separation between the $|s\rangle$ and $|p\rangle$ levels. Dashed lines: energies of non-interacting $|s,1 \rangle$ and $|p,0 \rangle$ level. (b) Scheme of the interaction between electronic levels and LO-phonon, leading to the polaron formation. Anharmonic coupling with the phonon reservoir and the far-infrared absorption are indicated (see text).}
\end{figure}
The two resulting polaron states ($i$=1,2) can be written:
\begin{equation}
|\psi_i\rangle = \alpha_i |p,0\rangle + \beta_i |s,1_{sp}\rangle
\end{equation}
The polaron energies $E_i$ are solutions of $(E_i-\hbar\omega_{\text{LO}})(E_i-E_{sp})=V_{sp}^2$, where $E_{sp}$ is the energy separation between the the $|s\rangle$ and $|p\rangle$ levels (experimentally, different $E_{sp}$ values correspond to QDs with different sizes and/or compositions).
The energy origin is taken at the $|s,0\rangle$ level
\footnote{We have neglected the small energy correction of the QD ground state due to non-resonant polaron couplings.},
so that the polaron energy $E_i$ corresponds to the one of the far-infrared photon (see Fig~1). The weight of the one-phonon components of the polaron states $i$ can be expressed as a function of the polaron energy $E_i$ as:
\begin{equation}
|\beta_i|^2 = \frac{V_{sp}^2}{ (E_{i}-\hbar\omega_{\text{LO}})^2 + V_{sp}^2}
\end{equation}

The anharmonic perturbation $V_a$ triggers the polaron relaxation. 
It couples the polaron state via its LO-phonon component to a reservoir of multi-phonon states.
Using Fermi golden rule, the broadenings of the polaron states are given by:

\begin{equation}
\Gamma_i (E_i) = |\beta_i|^2 ~ \Gamma^{\text{ph}} (E_i) \\
\end{equation}

\begin{equation}
\Gamma^{\text{ph}}(E_i) = 2\pi  \sum_{\nu} \vert\langle\nu|V_a|1_{sp}\rangle\vert^2 \delta(E_i - E_{\nu})
\end{equation}

where $\nu$ label the multi-phonons states into which the polaron disintegrate. The $|\beta_i|^2$ term follows from the fact that $V_a$ acts only on the LO-phonon part of the polaron eigenstate.
Note that the semi-classical approach leads to $\Gamma_i (E_i) = |\beta_i|^2\Gamma^{\text{ph}}(\hbar\omega_{\text{LO}}) $, where $\hbar/\Gamma^{\text{ph}}(\hbar\omega_{\text{LO}}) $ is the bulk phonon lifetime. In our model instead, the polaron lifetimes are evaluated at the polaron energies. Since the optically probed state has energy $E_i$ that can significantly differ from $\hbar\omega_{\text{LO}}$, the anharmonic coupling strength as well as the density of multi-phonons final states $|\nu\rangle$ resonant with $E_i$ may greatly differ from the bulk. Therefore, as we show below, $\Gamma^{\text{ph}}(E_i)$ can be very different from $\Gamma^{\text{ph}}(\hbar\omega_{\text{LO}})$.

\begin{figure}
\begin{center}
\includegraphics[width=0.45\textwidth]{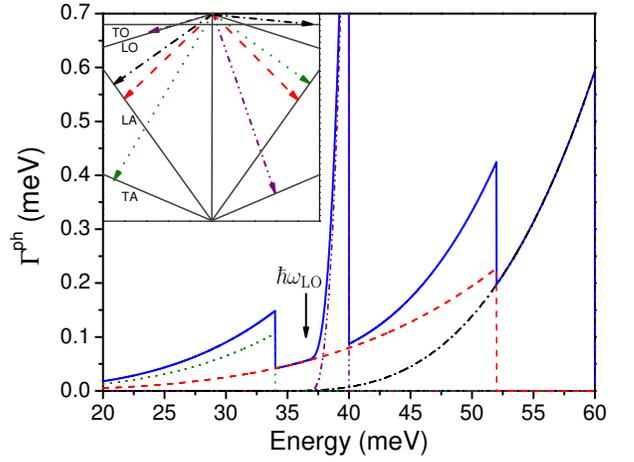}
\end{center}
\caption{\footnotesize (Color online) Calculated $\Gamma^{ph}$ versus the
energy:  Full calculation (solid line); LA+LA channel (Klemens channel) in dashed
line; LA+TA channel in dotted line; LA+TO channel in dashed-dotted line; TA+LO channel in dashed-double-dotted line (T=0K). Inset: Schematic of the different anharmonic processes. }
\label{gamma}
\end{figure}

Linewidth of bulk LO-phonon $\Gamma(\hbar\omega_{\text{LO}})$ has been calculated in several works: first principles calculations have been used by Debernardi \textit{et al.} \cite{debernardi95, debernardi98}, while Klemens \cite{klemens66} or more recently Barman and Srivastava \cite{barman04} have used a more  phenomenological approach. Here, we generalize the later model to study the relaxation of polarons in QDs.
We consider only the cubic part of the anharmonic vibrational hamiltonian \cite{barman04}:

\begin{equation}
\begin{split}
V_{a} = & \frac{\Omega}{3!\sqrt{N_{cr}}}\sum_{\V{k_0}j_0,\V{k_1}j_1,\V{k_2}j_2}
 P(\V{k_0},j_0;\V{k_1},j_1;\V{k_2},j_2) \\
& \times 
u_{\V{k_0},j_0}  u_{\V{k_1},j_1} u_{\V{k_2},j_2}
~ \delta_{\V{k_0}+\V{k_1}+\V{k_2}}
\\
\end{split}
\label{va}
\end{equation}

where $\Omega$ is the unit cell volume, $N_{cr}=V_{cr}/\Omega$ and
\begin{equation}
u_{\V{k},j}=\sqrt{\frac{\hbar}{2M\omega_{\V{k},j}}}(a_{\V{k},j} + a^+_{\V{-k},j})
\end{equation}

$V_a$ describes the coupling of phonon of branche $j_0$ with two-phonon states of branches $j_1$ and $j_2$. Each $(j_1,j_2)$ corresponds to a particular disintegration channel. The 3! factor accounts for equivalent processes obtained by cyclic permutation of the labels in the triple summation.

As the $\vert 1_{sp} \rangle$ LO-phonon mode is formed of bulk phonons with small wavevectors, momentum conservation in Eq.~\ref{va} implies a decay into a pair of phonons with opposite wavevectors.
Therefore, one gets approximatively at low temperature:
\begin{equation}
\begin{split}  
\Gamma^{\text{ph}}(E) = & \frac{\pi\hbar^3}{8V_{cr}\rho^3} \sum_{\V{k_1}j_1,\V{k_2}j_2}\delta_{\V{k_1}+\V{k_2}} \frac{\left\vert P(\V{0},LO;\V{k_1},j_1;\V{k_2},j_2)\right\vert ^2}{\omega_{\text{LO}} ~ \omega_{\V{k_1},j_1} ~ \omega_{\V{k_2},j_2}} \\
& \times \delta\left( {E-\hbar\omega_{\V{k_1},j_1}-\hbar\omega_{\V{k_2},j_2}} \right) \\
\end{split}
\end{equation}

Following reference \onlinecite{srivastava90}, which extents
Klemens' work \cite{klemens66}, we use:
\begin{equation}
\vert P(\V{k_0},j_0;\V{k_1},j_1;\V{k_2},j_2) \vert =  \frac{2 \rho \gamma}{\overline{c}}
\omega_{\V{k_0},j_0}\omega_{\V{k_1},j_1} \omega_{\V{k_2},j_2}
\label{approx}
\end{equation}
where $\gamma$ is the mode-averaged Gruneisen's constant (we use $\gamma=1.3$ in the following\cite{adachi94,wickboldt87}), $\rho$ is the mass density and $\overline{c}$ is the average acoustic phonon speed ($3/\overline{c}=1/c_{LA}+2/c_{TA}$).

$\Gamma^{\text{ph}}(E)$ can be decomposed into the contributions due to the different channels $(j_1,j_2)$:
\begin{equation}
\Gamma^{\text{ph}}(E) = \sum_{(j_1,j_2)} \Gamma_{( j_1,j_2)}^{\text{ph}}(E)
\end{equation}
Using Eq.~\ref{approx}, we can express each of these contributions as:
\begin{equation}
\begin{split}
\Gamma_{(j_1,j_2)}^{\text{ph}}(E)= & \frac{\pi\hbar^3 \gamma^2}{2V_{cr}\rho  \overline{c}^2 } \sum_{\V{k_1},\V{k_2}} \omega_{LO}\omega_{\V{k_1},j_1}\omega_{\V{k_2},j_2} \\
& \times  \delta_{\V{k_1}+\V{k_2}} ~ \delta\left( {E-\hbar\omega_{\V{k_1},j_1}-\hbar\omega_{\V{k_2},j_2}}\right)
\end{split}
\label{eqcont}
\end{equation}

As in Ref~\onlinecite{barman04}, we model acoustic modes with Debye's isotropic continuum scheme.
The linewidth induced by a  given channel ($j_1,j_2$) can be expressed as:
\begin{equation}
\Gamma_{(j_1,j_2)}^{\text{ph}}(E)=  \frac{\hbar^2\gamma^2}{4\pi\rho \overline{c}^2c_{j_1}^2(c_{j_1}+c_{j_2}) }  \omega_{LO} \omega_{j_1}^3\omega_{j_2}
\label{eqgc}
\end{equation}
where $c_j$ is the sound velocity of branch $j$ for acoustic branches ($c_{j_2}=\partial \omega_{j_2} / \partial k_{j_2} $ for optical branches). In Eq~\ref{eqgc}, $\omega_{j_1}$ and $\omega_{j_2}$ are the phonon frequencies that satisfy both deltas in Eq~\ref{eqcont}.
For a non-zero temperature, this linewidth has to be multiplied by $(1+n_{j_1})(1+n_{j_2})$, where $n_{j_i}$ is the Bose occupation factor at the energy $\hbar\omega_{j_i}$.
Figure~\ref{gamma} shows the calculated $\Gamma^{\text{ph}}$ versus the polaron energy, as well as the contributions to $\Gamma^{\text{ph}}$ from
different channels: $\Gamma_{(TA,LA)}^{\text{ph}}$, $\Gamma_{(LA,LA)}^{\text{ph}}$ (Klemens channel),  $\Gamma_{(TA,LO)}^{\text{ph}}$ (Vall\'ee-Bogani channel) and $\Gamma_{(LA,TO)}^{\text{ph}}$.
We obtain very strong variations of $\Gamma^{\text{ph}}$
within the range of polaron energies that can be obtained in
InAs/GaAs QDs.
This radically differs from the constant value $\Gamma^{\text{ph}}(\hbar\omega_{\text{LO}})$ of the semi-classical model. In order to understand the origin of these variations, let us consider the calculation of the Klemens channel :

\begin{equation}
\Gamma^{\text{ph}}_{\text{(LA,LA)}}(E)= \frac{\gamma^2}{128\pi\hbar^2\rho \overline{c}^2 c_{LA}^3 }  \omega_{LO}E^4
\end{equation}

The  $E^4$ dependence arises from two contributions: the density of final states with two LA phonons with energies
equal to half of the polaron one increases like $E^2$ (in the Debye
model we are considering), while the anharmonic coupling strength also increases
like $E^2$. The disintegration rate $\Gamma_{(LA,LA)}^{\text{ph}}(E)$ increases with $E$ until the energy of the emitted LA phonons reach their maximum value (corresponding to the emission of two zone-edge LA phonons, for $E \simeq 52$ meV).
We discuss further the implications of these results below. Let us first point out that we have checked that the Fermi golden rule can be safely applied in our case:
(i) the anharmonic self-energy\cite{debernardi99} (the Hilbert transform of  $\Gamma^{\text{ph}}(E)$) is negligible (on the 0.1 meV energy scale, except at channel edges) with respect to the typical variation of $E$ studied here.
(ii) $\text{d}\Gamma^{\text{ph}} / \text{d}E \ll 1$ except at channel edges, so that the time decay of the polaron population is exponential.
Finally, as shown in Fig~2, we find that the (LA,LA) channel is the dominant mechanism in bulk
\footnote{The dominant mechanism in bulk GaAs is a subject of debate\cite{vallee91,debernardi98,barman04}. This can be understood by noting that $\hbar\omega_{\text{LO}}$ is slightly above the high energy edge of the (TA,LA) and slightly below the rise of (TA,LO), and thus the disintegration in bulk depends on the details of the phonon dispersion near the zone edges.}.

\begin{figure}
\begin{center}
\includegraphics[width=.45\textwidth]{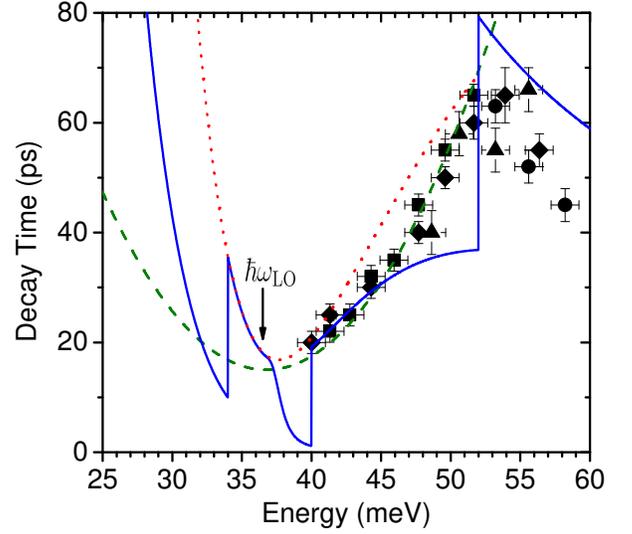}
\end{center}
\caption{\footnotesize (Color online) Polaron decay time as a function of the polaron energy: measurements from Ref~\onlinecite{zibik04} in symbols, full calculation (presented in the text) in solid line, calculation involving only the Klemens channel (LA+LA) in dotted line and best fit obtained with the semi-classical model in dashed line ($V_{sp\text{(fit)}}=8$meV, $\tau_{LO}=15$ps).}
\label{polaron}
\end{figure}

The final polaron lifetime $\tau_{pol}=\hbar/\Gamma_i$ (including both the results in Fig~2 and the energy-dependent weights $|\beta_i|^2$) as a function of polaron energy $E_{pol}=E_{i=1,2}$ is shown in Fig~3 (solid lines), and compared to the data of Ref~\onlinecite{zibik04} (symbols).  We observe a good overall agreement with the measurements for any polaron energy.  Before discussing the complex energy dependence of $\tau_{pol}$, let us consider the applicability of the semi-classical model.  Firstly, the best fit using this model (dashed line in Fig~3) is obtained for $V_{sp(fit)} \simeq 8$ meV, which is about twice the calculated value of the Fr\"ohlich coupling strength for the QDs studied in Refs~\onlinecite{sauvage02} and \onlinecite{zibik04}.  Note that $V_{sp} \simeq 4$ meV is found to be roughly independent of the polaron energy, so that the factor of two in $V_{sp\text{(fit)}}$ cannot result from the inhomogeneous distribution of QDs.  Using instead the calculated value of $V_{sp}$ in the semi-classical formula would lead to a much steeper increase in Fig~3, since $\tau_{pol} \simeq \hbar(E_{pol}-\hbar\omega_{LO})^2/[\Gamma^{\text{ph}}(\omega_{LO}) V_{sp}^2] $ for high energy polarons
\footnote{Note that in the work of Li \textit{et al}\cite{li99}, $\Gamma$ in formula (10) does not represent the bulk decay rate but half this value, as follows from the definition of $\Gamma_k$ in formula (3).  Also, at resonance ($\Delta=0$ in their work) the polaron lifetime should be twice longer than the LO phonon one, because of its one-phonon weight $|\beta|^2 = 1/2$.  This missing factor of two affects the value of $V_{sp\text{(fit)}}$ extracted from a best-fit procedure (see \onlinecite{sauvage02}, \onlinecite{zibik04}).}.
 Secondly, the semi-classical model predicts a continuous increase of $\tau_{pol}$ with the absolute value of the energy detuning $|E_{pol}-\hbar\omega_{LO}|$. Our results do not support this prediction, as discussed in the following.

Indeed, the existence of different channels, with specific energy windows, leads to important variations of $\tau_{pol}$ with $E_{pol}$, in particular near the windows edges.  We associate the pronounced discontinuity in the data of Ref~\onlinecite{zibik04} at $\simeq 52$ meV with the high-energy limit of the Klemens channel.  According to Fig~2, this disintegration channel dominates over a large energy interval.  This is in agreement with other data of Zibik \textit{et al}, who measured the temperature dependence of the polaron lifetime at $E_{pol}$ = 44 meV and found that the dominant mechanism at this energy was the (LA,LA) one \cite{zibik04}.  Furthermore, we clearly see from Figs~2 and 3 that the dominant disintegration path depends on the polaron energy.  We associate the important decrease of $\tau_{pol}$ above 52 meV with the (LA,TO) channel.  It is worth stressing that the efficiency of the anharmonic disintegration increases in this region despite the fact that the weight of the one-phonon component continuously decreases with increasing E.  This counterintuitive result highlights again the importance of employing an accurate description of the anharmonic coupling and of the two-phonon density of states.  Finally, we predict that the disintegration of low energy polarons ($E_1 \ll \hbar\omega_{LO}$ in Fig~1a) should be governed by the (LA,TA) channel.  Note that lower branch polarons ($|\psi_1\rangle$ in Fig~1a) with energies near (but below) $\hbar\omega_{LO}$ have lower lifetime than the upper branch ones, and also that $\tau_{pol}$ increases much faster with increasing detuning in the lower branch than in the upper one.  
Let us finally stress that we do not have any fit parameter in our model, but consider bulk-like values for the different material parameters, in particular an averaged channel-independent value for $\gamma$.  This approximation, together with the very simple (isotropic and linear) phonon dispersion we consider, may explain the discrepancies between calculated and measured lifetimes, especially near the zone edges.

In conclusion, we have presented in this work a complete study of polaron relaxation in self-assembled QDs.  We have shown that the decay rate of a QD polaron is equal to the product of its phonon weight by the decay rate of a zone-center bulk LO phonon \textit{which would have the polaron energy}.  The good overall agreement of our calculations with reported data on the population lifetime of excited QD states strongly supports the mandatory use of an accurate description of phonon anharmonicity within the polaron framework to describe the relaxation process in these structures. Moreover, we have presented a critical discussion of the previous models proposed to tackle the polaron relaxation, in particular the often quoted semi-classical one.  Finally, we hope that our work will stimulate further experiments, in particular for low-energy polarons pertaining to the lower branch (see Fig~1a), which are expected to be less sensitive to the anharmonic disintegration.  It would also be interesting to study higher energy states, for which higher order anharmonic terms may become important.

\acknowledgments
The LPA-ENS (UMR 8551) is associated with the CNRS and the Universit\'es Paris~6 and Paris~7. We thank Drs. E. A. Zibik, L. R. Wilson and Prof. M . S. Skolnick  for very helpful discussions.


\begin{thebibliography}{24}
\expandafter\ifx\csname natexlab\endcsname\relax\def\natexlab#1{#1}\fi
\expandafter\ifx\csname bibnamefont\endcsname\relax
  \def\bibnamefont#1{#1}\fi
\expandafter\ifx\csname bibfnamefont\endcsname\relax
  \def\bibfnamefont#1{#1}\fi
\expandafter\ifx\csname citenamefont\endcsname\relax
  \def\citenamefont#1{#1}\fi
\providecommand{\bibinfo}[2]{#2}
\providecommand{\eprint}[2][]{\url{#2}}

\bibitem[{\citenamefont{Bockelmann and Bastard}(1990)}]{bockelmann90}
\bibinfo{author}{\bibfnamefont{U.}~\bibnamefont{Bockelmann}} \bibnamefont{and}
  \bibinfo{author}{\bibfnamefont{G.}~\bibnamefont{Bastard}},
  \bibinfo{journal}{Phys. Rev. B} \textbf{\bibinfo{volume}{42}},
  \bibinfo{pages}{8947} (\bibinfo{year}{1990}).

\bibitem[{\citenamefont{Benisty et~al.}(1991)\citenamefont{Benisty,
  Sotomayor-Torr\`es, and Weisbuch}}]{benisty91}
\bibinfo{author}{\bibfnamefont{H.}~\bibnamefont{Benisty}},
  \bibinfo{author}{\bibfnamefont{C.~M.} \bibnamefont{Sotomayor-Torr\`es}},
  \bibnamefont{and} \bibinfo{author}{\bibfnamefont{C.}~\bibnamefont{Weisbuch}},
  \bibinfo{journal}{Phys. Rev. B} \textbf{\bibinfo{volume}{44}},
  \bibinfo{pages}{10945} (\bibinfo{year}{1991}).

\bibitem[{\citenamefont{Inoshita and Sakaki}(1992)}]{inoshita92}
\bibinfo{author}{\bibfnamefont{T.}~\bibnamefont{Inoshita}} \bibnamefont{and}
  \bibinfo{author}{\bibfnamefont{H.}~\bibnamefont{Sakaki}},
  \bibinfo{journal}{Phys. Rev. B} \textbf{\bibinfo{volume}{46}},
  \bibinfo{pages}{7260} (\bibinfo{year}{1992}).

\bibitem[{\citenamefont{Inoshita and Sakaki}(1997)}]{inoshita97}
\bibinfo{author}{\bibfnamefont{T.}~\bibnamefont{Inoshita}} \bibnamefont{and}
  \bibinfo{author}{\bibfnamefont{H.}~\bibnamefont{Sakaki}},
  \bibinfo{journal}{Phys. Rev. B} \textbf{\bibinfo{volume}{56}},
  \bibinfo{pages}{R4355} (\bibinfo{year}{1997}).

\bibitem[{\citenamefont{Hameau et~al.}(1999)\citenamefont{Hameau, Guldner,
  Verzelen, Ferreira, Bastard, Zeman, Lema\^itre, and G\'erard}}]{hameau99}
\bibinfo{author}{\bibfnamefont{S.}~\bibnamefont{Hameau}},
  \bibinfo{author}{\bibfnamefont{Y.}~\bibnamefont{Guldner}},
  \bibinfo{author}{\bibfnamefont{O.}~\bibnamefont{Verzelen}},
  \bibinfo{author}{\bibfnamefont{R.}~\bibnamefont{Ferreira}},
  \bibinfo{author}{\bibfnamefont{G.}~\bibnamefont{Bastard}},
  \bibinfo{author}{\bibfnamefont{J.}~\bibnamefont{Zeman}},
  \bibinfo{author}{\bibfnamefont{A.}~\bibnamefont{Lema\^itre}},
  \bibnamefont{and} \bibinfo{author}{\bibfnamefont{J.~M.}
  \bibnamefont{G\'erard}}, \bibinfo{journal}{Phys. Rev. Lett.}
  \textbf{\bibinfo{volume}{83}}, \bibinfo{pages}{4152} (\bibinfo{year}{1999}).

\bibitem[{\citenamefont{Hameau et~al.}(2002)\citenamefont{Hameau, Isaia,
  Guldner, Deleporte, Verzelen, Ferreira, Bastard, Zeman, and
  G\'erard}}]{hameau02}
\bibinfo{author}{\bibfnamefont{S.}~\bibnamefont{Hameau}},
  \bibinfo{author}{\bibfnamefont{J.~N.} \bibnamefont{Isaia}},
  \bibinfo{author}{\bibfnamefont{Y.}~\bibnamefont{Guldner}},
  \bibinfo{author}{\bibfnamefont{E.}~\bibnamefont{Deleporte}},
  \bibinfo{author}{\bibfnamefont{O.}~\bibnamefont{Verzelen}},
  \bibinfo{author}{\bibfnamefont{R.}~\bibnamefont{Ferreira}},
  \bibinfo{author}{\bibfnamefont{G.}~\bibnamefont{Bastard}},
  \bibinfo{author}{\bibfnamefont{J.}~\bibnamefont{Zeman}}, \bibnamefont{and}
  \bibinfo{author}{\bibfnamefont{J.~M.} \bibnamefont{G\'erard}},
  \bibinfo{journal}{Phys. Rev. B} \textbf{\bibinfo{volume}{65}},
  \bibinfo{pages}{085316} (\bibinfo{year}{2002}).

\bibitem[{\citenamefont{Carpenter et~al.}(2006)\citenamefont{Carpenter, Zibik,
  Sadowski, Wilson, Whittaker, Cockburn, Skolnick, Potemski, Steer, and
  Hopkinson}}]{carpenter06}
\bibinfo{author}{\bibfnamefont{B. A.}~\bibnamefont{Carpenter}},
  \bibinfo{author}{\bibfnamefont{E. A.}~\bibnamefont{Zibik}},
  \bibinfo{author}{\bibfnamefont{M. L.}~\bibnamefont{Sadowski}},
  \bibinfo{author}{\bibfnamefont{L. R.}~\bibnamefont{Wilson}},
  \bibinfo{author}{\bibfnamefont{D. M.}~\bibnamefont{Whittaker}},
  \bibinfo{author}{\bibfnamefont{J. W.}~\bibnamefont{Cockburn}},
  \bibinfo{author}{\bibfnamefont{M. S.}~\bibnamefont{Skolnick}},
  \bibinfo{author}{\bibfnamefont{M.}~\bibnamefont{Potemski}},
  \bibinfo{author}{\bibfnamefont{M. J.}~\bibnamefont{Steer}}, \bibnamefont{and}
  \bibinfo{author}{\bibfnamefont{M.}~\bibnamefont{Hopkinson}},
  \bibinfo{journal}{Phys. Rev. B} \textbf{\bibinfo{volume}{74}},
  \bibinfo{pages}{161302(R)} (\bibinfo{year}{2006}).

\bibitem[{\citenamefont{Preisler et~al.}(2005)\citenamefont{Preisler, Ferreira,
  Hameau, de~Vaulchier, Guldner, Sadowski, and Lemaitre}}]{preisler05}
\bibinfo{author}{\bibfnamefont{V.}~\bibnamefont{Preisler}},
  \bibinfo{author}{\bibfnamefont{R.}~\bibnamefont{Ferreira}},
  \bibinfo{author}{\bibfnamefont{S.}~\bibnamefont{Hameau}},
  \bibinfo{author}{\bibfnamefont{L. A.}~\bibnamefont{de~Vaulchier}},
  \bibinfo{author}{\bibfnamefont{Y.}~\bibnamefont{Guldner}},
  \bibinfo{author}{\bibfnamefont{M. L.}~\bibnamefont{Sadowski}}, \bibnamefont{and}
  \bibinfo{author}{\bibfnamefont{A.}~\bibnamefont{Lemaitre}},
  \bibinfo{journal}{Phys. Rev. B} \textbf{\bibinfo{volume}{72}},
  \bibinfo{pages}{115309} (\bibinfo{year}{2005}).

\bibitem[{\citenamefont{Preisler et~al.}(2006)\citenamefont{Preisler, Grange,
  Ferreira, de~Vaulchier, Guldner, Teran, Potemski, and
  Lema{\^\i}tre}}]{preisler06}
\bibinfo{author}{\bibfnamefont{V.}~\bibnamefont{Preisler}},
  \bibinfo{author}{\bibfnamefont{T.}~\bibnamefont{Grange}},
  \bibinfo{author}{\bibfnamefont{R.}~\bibnamefont{Ferreira}},
  \bibinfo{author}{\bibfnamefont{L. A.}~\bibnamefont{de~Vaulchier}},
  \bibinfo{author}{\bibfnamefont{Y.}~\bibnamefont{Guldner}},
  \bibinfo{author}{\bibfnamefont{F. J.}~\bibnamefont{Teran}},
  \bibinfo{author}{\bibfnamefont{M.}~\bibnamefont{Potemski}}, \bibnamefont{and}
  \bibinfo{author}{\bibfnamefont{A.}~\bibnamefont{Lemaitre}},
  \bibinfo{journal}{Phys. Rev. B} \textbf{\bibinfo{volume}{73}},
  \bibinfo{pages}{75320} (\bibinfo{year}{2006}).

\bibitem[{\citenamefont{Sauvage et~al.}(2002)\citenamefont{Sauvage, Boucaud,
  Lobo, Bras, Fishman, Prazeres, Glotin, Ortega, and G\'erard}}]{sauvage02}
\bibinfo{author}{\bibfnamefont{S.}~\bibnamefont{Sauvage}},
  \bibinfo{author}{\bibfnamefont{P.}~\bibnamefont{Boucaud}},
  \bibinfo{author}{\bibfnamefont{R.~P. S.~M.} \bibnamefont{Lobo}},
  \bibinfo{author}{\bibfnamefont{F.}~\bibnamefont{Bras}},
  \bibinfo{author}{\bibfnamefont{G.}~\bibnamefont{Fishman}},
  \bibinfo{author}{\bibfnamefont{R.}~\bibnamefont{Prazeres}},
  \bibinfo{author}{\bibfnamefont{F.}~\bibnamefont{Glotin}},
  \bibinfo{author}{\bibfnamefont{J.~M.} \bibnamefont{Ortega}},
  \bibnamefont{and} \bibinfo{author}{\bibfnamefont{J.-M.}
  \bibnamefont{G\'erard}}, \bibinfo{journal}{Phys. Rev. Lett.}
  \textbf{\bibinfo{volume}{88}}, \bibinfo{pages}{177402}
  (\bibinfo{year}{2002}).

\bibitem[{\citenamefont{Zibik et~al.}(2004)\citenamefont{Zibik, Wilson, Green,
  Bastard, Ferreira, Phillips, Carder, Wells, Cockburn, Skolnick
  et~al.}}]{zibik04}
\bibinfo{author}{\bibfnamefont{E.~A.} \bibnamefont{Zibik}},
  \bibinfo{author}{\bibfnamefont{L.~R.} \bibnamefont{Wilson}},
  \bibinfo{author}{\bibfnamefont{R.~P.} \bibnamefont{Green}},
  \bibinfo{author}{\bibfnamefont{G.}~\bibnamefont{Bastard}},
  \bibinfo{author}{\bibfnamefont{R.}~\bibnamefont{Ferreira}},
  \bibinfo{author}{\bibfnamefont{P.~J.} \bibnamefont{Phillips}},
  \bibinfo{author}{\bibfnamefont{D.~A.} \bibnamefont{Carder}},
  \bibinfo{author}{\bibfnamefont{J.-P.~R.} \bibnamefont{Wells}},
  \bibinfo{author}{\bibfnamefont{J.~W.} \bibnamefont{Cockburn}},
  \bibinfo{author}{\bibfnamefont{M.~S.} \bibnamefont{Skolnick}},
  \bibinfo{author}{\bibfnamefont{M.~J.} \bibnamefont{Steer}},
  \bibinfo{author}{\bibfnamefont{M.} \bibnamefont{Hopkinson}},
  \bibnamefont{et~al.}, \bibinfo{journal}{Physical Review B}
  \textbf{\bibinfo{volume}{70}}, \bibinfo{pages}{161305(R)}
  (\bibinfo{year}{2004}).

\bibitem[{\citenamefont{Grange et~al.}(2007)\citenamefont{Grange, Zibik,
  Ferreira, Bastard, Carpenter, Phillips, Stehr, Winnerl, Helm, Steer
  et~al.}}]{grange07}
\bibinfo{author}{\bibfnamefont{T.}~\bibnamefont{Grange}},
  \bibinfo{author}{\bibfnamefont{E.}~\bibnamefont{Zibik}},
  \bibinfo{author}{\bibfnamefont{R.}~\bibnamefont{Ferreira}},
  \bibinfo{author}{\bibfnamefont{G.}~\bibnamefont{Bastard}},
  \bibinfo{author}{\bibfnamefont{B.}~\bibnamefont{Carpenter}},
  \bibinfo{author}{\bibfnamefont{P.}~\bibnamefont{Phillips}},
  \bibinfo{author}{\bibfnamefont{D.}~\bibnamefont{Stehr}},
  \bibinfo{author}{\bibfnamefont{S.}~\bibnamefont{Winnerl}},
  \bibinfo{author}{\bibfnamefont{M.}~\bibnamefont{Helm}},
  \bibinfo{author}{\bibfnamefont{M.}~\bibnamefont{Steer}},
  \bibnamefont{et~al.}, \bibinfo{journal}{New J. Phys.}
  \textbf{\bibinfo{volume}{9}}, \bibinfo{pages}{259} (\bibinfo{year}{2007}).

\bibitem[{\citenamefont{Li et~al.}(1999)\citenamefont{Li, Nakayama, and
  Arakawa}}]{li99}
\bibinfo{author}{\bibfnamefont{X.-Q.} \bibnamefont{Li}},
  \bibinfo{author}{\bibfnamefont{H.}~\bibnamefont{Nakayama}}, \bibnamefont{and}
  \bibinfo{author}{\bibfnamefont{Y.}~\bibnamefont{Arakawa}},
  \bibinfo{journal}{Phys. Rev. B} \textbf{\bibinfo{volume}{59}},
  \bibinfo{pages}{5069} (\bibinfo{year}{1999}).

\bibitem[{\citenamefont{Verzelen et~al.}(2000)\citenamefont{Verzelen, Ferreira,
  and Bastard}}]{verzelen00}
\bibinfo{author}{\bibfnamefont{O.}~\bibnamefont{Verzelen}},
  \bibinfo{author}{\bibfnamefont{R.}~\bibnamefont{Ferreira}}, \bibnamefont{and}
  \bibinfo{author}{\bibfnamefont{G.}~\bibnamefont{Bastard}},
  \bibinfo{journal}{Phys. Rev. B} \textbf{\bibinfo{volume}{62}},
  \bibinfo{pages}{R4809} (\bibinfo{year}{2000}).

\bibitem[{\citenamefont{Jacak et~al.}(2002)\citenamefont{Jacak, Krasnyj, Jacak,
  and Machnikowski}}]{jacak02}
\bibinfo{author}{\bibfnamefont{L.}~\bibnamefont{Jacak}},
  \bibinfo{author}{\bibfnamefont{J.}~\bibnamefont{Krasnyj}},
  \bibinfo{author}{\bibfnamefont{D.}~\bibnamefont{Jacak}}, \bibnamefont{and}
  \bibinfo{author}{\bibfnamefont{P.}~\bibnamefont{Machnikowski}},
  \bibinfo{journal}{Phys. Rev. B} \textbf{\bibinfo{volume}{65}},
  \bibinfo{pages}{113305} (\bibinfo{year}{2002}).

\bibitem[{\citenamefont{Vall\'ee and Bogani}(1991)}]{vallee91}
\bibinfo{author}{\bibfnamefont{F.}~\bibnamefont{Vall\'ee}} \bibnamefont{and}
  \bibinfo{author}{\bibfnamefont{F.}~\bibnamefont{Bogani}},
  \bibinfo{journal}{Phys. Rev. B} \textbf{\bibinfo{volume}{43}},
  \bibinfo{pages}{12049} (\bibinfo{year}{1991}).

\bibitem[{\citenamefont{Debernardi et~al.}(1995)\citenamefont{Debernardi,
  Baroni, and Molinari}}]{debernardi95}
\bibinfo{author}{\bibfnamefont{A.}~\bibnamefont{Debernardi}},
  \bibinfo{author}{\bibfnamefont{S.}~\bibnamefont{Baroni}}, \bibnamefont{and}
  \bibinfo{author}{\bibfnamefont{E.}~\bibnamefont{Molinari}},
  \bibinfo{journal}{Phys. Rev. Lett.} \textbf{\bibinfo{volume}{75}},
  \bibinfo{pages}{1819} (\bibinfo{year}{1995}).

\bibitem[{\citenamefont{Debernardi}(1998)}]{debernardi98}
\bibinfo{author}{\bibfnamefont{A.}~\bibnamefont{Debernardi}},
  \bibinfo{journal}{Phys. Rev. B} \textbf{\bibinfo{volume}{57}},
  \bibinfo{pages}{12847} (\bibinfo{year}{1998}).

\bibitem[{\citenamefont{Klemens}(1966)}]{klemens66}
\bibinfo{author}{\bibfnamefont{P.~G.} \bibnamefont{Klemens}},
  \bibinfo{journal}{Phys. Rev.} \textbf{\bibinfo{volume}{148}},
  \bibinfo{pages}{845} (\bibinfo{year}{1966}).

\bibitem[{\citenamefont{Barman and Srivastava}(2004)}]{barman04}
\bibinfo{author}{\bibfnamefont{S.}~\bibnamefont{Barman}} \bibnamefont{and}
  \bibinfo{author}{\bibfnamefont{G. P.}~\bibnamefont{Srivastava}},
  \bibinfo{journal}{Phys. Rev. B} \textbf{\bibinfo{volume}{69}},
  \bibinfo{pages}{235208} (\bibinfo{year}{2004}).

\bibitem[{\citenamefont{Srivastava}(1990)}]{srivastava90}
\bibinfo{author}{\bibfnamefont{G.}~\bibnamefont{Srivastava}},
  \emph{\bibinfo{title}{{The Physics of Phonons}}} (\bibinfo{publisher}{Hilger,
  Bristol}, \bibinfo{year}{1990}).

\bibitem[{\citenamefont{Adachi}(1994)}]{adachi94}
\bibinfo{author}{\bibfnamefont{S.}~\bibnamefont{Adachi}},
  \emph{\bibinfo{title}{{GaAs and Related Materials: Bulk Semiconducting and
  Superlattice Properties}}} (\bibinfo{publisher}{World Scientific},
  \bibinfo{year}{1994}).

\bibitem[{\citenamefont{Wickboldt et~al.}(1987)\citenamefont{Wickboldt,
  Anastassakis, Sauer, and Cardona}}]{wickboldt87}
\bibinfo{author}{\bibfnamefont{P.}~\bibnamefont{Wickboldt}},
  \bibinfo{author}{\bibfnamefont{E.}~\bibnamefont{Anastassakis}},
  \bibinfo{author}{\bibfnamefont{R.}~\bibnamefont{Sauer}}, \bibnamefont{and}
  \bibinfo{author}{\bibfnamefont{M.}~\bibnamefont{Cardona}},
  \bibinfo{journal}{Phys. Rev. B} \textbf{\bibinfo{volume}{35}},
  \bibinfo{pages}{1362} (\bibinfo{year}{1987}).

\bibitem[{\citenamefont{Debernardi}(1999)}]{debernardi99}
\bibinfo{author}{\bibfnamefont{A.}~\bibnamefont{Debernardi}},
  \bibinfo{journal}{Solid State Commun.} \textbf{\bibinfo{volume}{113}},
  \bibinfo{pages}{1} (\bibinfo{year}{1999}).

\end{thebibliography}

\end{document}